\documentclass[twocolumn,aps,prl,amsmath,superscriptaddress]{revtex4}
        \usepackage{amsmath}
        \usepackage{amssymb}
        \usepackage{natbib}
        \usepackage{epsfig}
        \usepackage{amsfonts}
        \usepackage{mathrsfs}
        \usepackage{color}
        \usepackage{bm}
        \usepackage[toc,page,title,titletoc,header]{appendix}
        \usepackage{CJK}
        \usepackage{epstopdf}
    \usepackage{mathrsfs}
    \usepackage{float}
    \usepackage{tikz-cd}
    \usepackage{graphicx}
    \usepackage{dcolumn}
    \usepackage{bm}
\usepackage{multirow}
\usepackage{comment}
\usepackage{indentfirst}
   \begin{document}
      \widetext
     \title{Helical damping and anomalous critical non-Hermitian skin effect}
      \author{Chun-Hui Liu}
       \affiliation{Beijing National Laboratory for Condensed Matter Physics, Institute of Physics, Chinese Academy of Sciences, Beijing 100190, China}
\affiliation{School of Physical Sciences, University of Chinese Academy of Sciences, Beijing 100049, China}
      \author{Kai Zhang}
       \affiliation{Beijing National Laboratory for Condensed Matter Physics, Institute of Physics, Chinese Academy of Sciences, Beijing 100190, China}
\affiliation{School of Physical Sciences, University of Chinese Academy of Sciences, Beijing 100049, China}
      \author{Zhesen Yang}
      \affiliation{Beijing National Laboratory for Condensed Matter Physics, Institute of Physics, Chinese Academy of Sciences, Beijing 100190, China}
\affiliation{School of Physical Sciences, University of Chinese Academy of Sciences, Beijing 100049, China}
      \author{Shu Chen}
      \email{schen@iphy.ac.cn}
\affiliation{Beijing National Laboratory for Condensed Matter Physics, Institute of Physics, Chinese Academy of Sciences, Beijing 100190, China}
\affiliation{School of Physical Sciences, University of Chinese Academy of Sciences, Beijing 100049, China}
\affiliation{Yangtze River Delta Physics Research Center, Liyang, Jiangsu 213300, China}
          \begin{abstract}
           \par
           Non-Hermitian skin effect and critical skin effect are  unique features of non-Hermitian systems.
           In this Letter, we study an open system with its dynamics of single-particle correlation function effectively dominated by a non-Hermitian damping matrix, which exhibits $\mathbb{Z}_2$ skin effect, and uncover the existence of a novel phenomenon of helical damping. When adding perturbations that break anomalous time reversal symmetry to the system, the critical skin effect occurs, which causes the disappearance of the helical damping in the thermodynamic limit although it can exist in small size systems.
           We also demonstrate the existence of anomalous critical skin effect when we couple two identical systems with $\mathbb{Z}_2$ skin effect. With the help of non-Bloch band theory, we unveil that the change of generalized Brillouin zone equation is the necessary condition of critical skin effect.

          \end{abstract}
          \maketitle
          \emph{Introduction.}--- Research on non-Hermitian systems is attracting growing attention as they demonstrate some novel properties without Hermitian counterparts \cite{TELee,Leykam,ShenH,Gong,XuY,Yin,Alvarez,Lieu,SYao1,SYao2,FSong2,CHLiu1,CHLiu2,NHatano,TYoshida,Xiong,Kunst,Sato,Zhou,Esaki,Runder} and many physical problems in photonic systems, electrical systems and open quantum systems  can be converted to non-Hermitian Hamiltonian problems \cite{Ozawa,Guo,Peng,Zoller,Carmichael,Rotter,Harari}. In general, a Markovian open quantum system can be mapped to the problem of density matrix evolution in terms of the Lindblad equation \cite{GLindblad,GKS}.
          If the Hamiltonian of the system is quadratic and the Lindblad operators are linear, the solution of Lindblad equation can be reduced to solving quadratic non-Hermitian Liouvillian matrix \cite{TProsen1,TProsen2}.
          While topological edge states of non-Hermitian Hamiltonians have been intensively studied in recent years \cite{Slager,Kou,ZhuBG,Nori,Menke,Yuce,Wunner,ChenR}, 
          it is insufficient to study the unique features of non-Hermitian matrix in open quantum systems \cite{TProsen3,SDiehl,CEBardyn,SLieu,FSong1,Goldstein,Caspel,Runder2019}.

          One of unique features of non-Hermitian systems is the non-Hermitian skin effect \cite{SYao1}, which is characterized by the emergence of some eigenstates which corresponding to bulk continuous eigenvalues localized at the boundaries, accompanied with the inconformity  of the open and periodic boundary energy spectrum,
          and breakdown of conventional bulk boundary correspondence \cite{SYao1,JiangH,LeeCH,LeeCH2,Xiong,Kunst,Ghatak,PXue,Longi,Helbig,YiW,Herviou,JinL,Kunst2019,LeeCH3}. Both phenomena can be understood in the scheme of non-Bloch band theory by introducing the concept of generalized Brillouin zone (GBZ). The GBZ is composed of all possible values of $z=e^{i(k+i\kappa)}$, where $k+i\kappa$ is the complex analytical continuation of Bloch momentum $k$,  and $\kappa$ is a function of $k$ and band index. The complex number $z$ can be derived  from the characteristic equation $f(z,E) \equiv det(H(z)-E) = 0$. By requiring a pair of zeros of the polynomial $f(z,E)$ to fulfill GBZ equation $|z_\mu|=|z_\nu|$ for the same $E$ and certain $\mu,\nu$, the GBZ of the system can be determined \cite{KYokomizo,KZhang,YYi,ZYang}. For systems with different symmetries, we note that equations for determining the GBZ may be different. By replacing BZ with GBZ both the bulk wave functions and eigenvalues of open boundary systems can be restored.
          Meanwhile, the skin effect is also unveiled to be originated from intrinsic non-Hermitian topology, which can be enriched by symmetry.
          This leads to the discovery of $\mathbb{Z}$ and $\mathbb{Z}_2$ non-Hermitian skin effect \cite{KZhang, NOkuma}. For open quantum systems related to non-Hermitian Hamiltonian with skin effect, the chiral damping has been uncovered \cite{FSong1}.

          The critical skin effect (CSE) is a rather unique phenomenon of the non-Hermitian system without Hermitian analogy. Very recently, CSE was dubbed to describe a novel critical behavior in the non-Hermitian system with the energy spectrum and wave function jumping discontinuously across a critical point \cite{LLi}.  It is revealed by ref \cite{LLi} that CSE occurs whenever one band subsystems with different GBZs are coupled by even a vanishingly small $k$ independent perturbation.
          According to Ref. \cite{LLi}, CSE does not occur when two one-band subsystems with the same GBZ are coupled by $k$ independent perturbation. We construct an example with CSE by using perturbation to couple systems with same GBZs, for which we call it {\it anomalous critical skin effect}. And we also construct an example that  subsystems with different GBZs are coupled by perturbations but don't support CSE. We shall explain these phenomena and demonstrate that 
          {\it the change of GBZ equation is the necessary condition of CSE. }

          In this paper, we shall work in open quantum systems described by Lindblad equation as our another important motivation is to explore new physical phenomenon associated with the $\mathbb{Z}_2$ skin effect and CSE in open quantum systems.  We consider a system with internal spin degree and demonstrate the existence of helical damping related to $\mathbb{Z}_2$ skin effect. The {\it helical damping} is characterized by the evolution of relative particle number $\tilde n(x,t)$ with exponentially decreasing intervals and power decreasing intervals distinguished by sharp wave fronts with opposite propagation directions. When the coupling perturbation breaks the anomalous time reversal symmetry, we demonstrate that the corresponding damping matrix exhibits CSE which leads to the disappearance of helical damping under the thermodynamic limit. Our research provides a framework for studying CSE and symmetry protected skin effect in open quantum systems and reveal the origin of CSE.

     \emph{Helical damping.}--- Open Markovian quantum systems satisfy the Lindblad master equation \cite{GLindblad,GKS}:
     \begin{equation}
    \frac{d\rho}{dt} 
    =-i[H,\rho]+\sum _{\mu}(2L_{\mu}\rho L_{\mu}^{\dagger}-\left\{L_{\mu}^{\dagger}L_{\mu},\rho\right\}) ,
    \label{eq1}
    \end{equation}
    where $\rho$ is the density matrix, $H$ is the Hamiltonian and $L_{\mu}$ are the Lindblad operators describing quantum jumps induced by the coupling to the environment.
          Consider a one-dimensional (1D) lattice with the unit cell composed of two orbits (sublattices) and each site can be occupied by spin up and spin down fermions. In the momentum space, the Hamiltonian is given by
    \begin{equation}
    h(k)=t_1\sigma_x+(t_2\sigma_y+\delta_1\tau_x) \sin k+t_2\sigma_x \cos k+\delta_2\sigma_z\tau_x, \label{h}
    \end{equation}
   where $\sigma_{x,y,z}$ and $\tau_{x,y,z}$ act on orbit and spin degree of freedom, respectively. Here we consider quantum jump processes described by the following Lindblad operators:
   \begin{equation}
   \begin{split}
    L_{x\uparrow}^l=\sqrt{\frac{\gamma_l}{2}}(c_{xA\uparrow}-ic_{xB\uparrow}), \quad L_{x\uparrow}^g=\sqrt{\frac{\gamma_g}{2}}(c^{\dagger}_{xA\uparrow}+ic^{\dagger}_{xB\uparrow}), \\
    L_{x\downarrow}^l=\sqrt{\frac{\gamma_l}{2}}(c_{xA\downarrow}+ic_{xB\downarrow}), \quad L_{x\downarrow}^g=\sqrt{\frac{\gamma_g}{2}}(c^{\dagger}_{xA\downarrow}-ic^{\dagger}_{xB\downarrow});
    \end{split}
    \label{operators}
    \end{equation}
where $s=\uparrow, \downarrow$ and $o=A, B$ refer to the spin and orbit index, respectively. And $x$ is cell index.

\begin{figure} 
\includegraphics[width=1.0\linewidth]{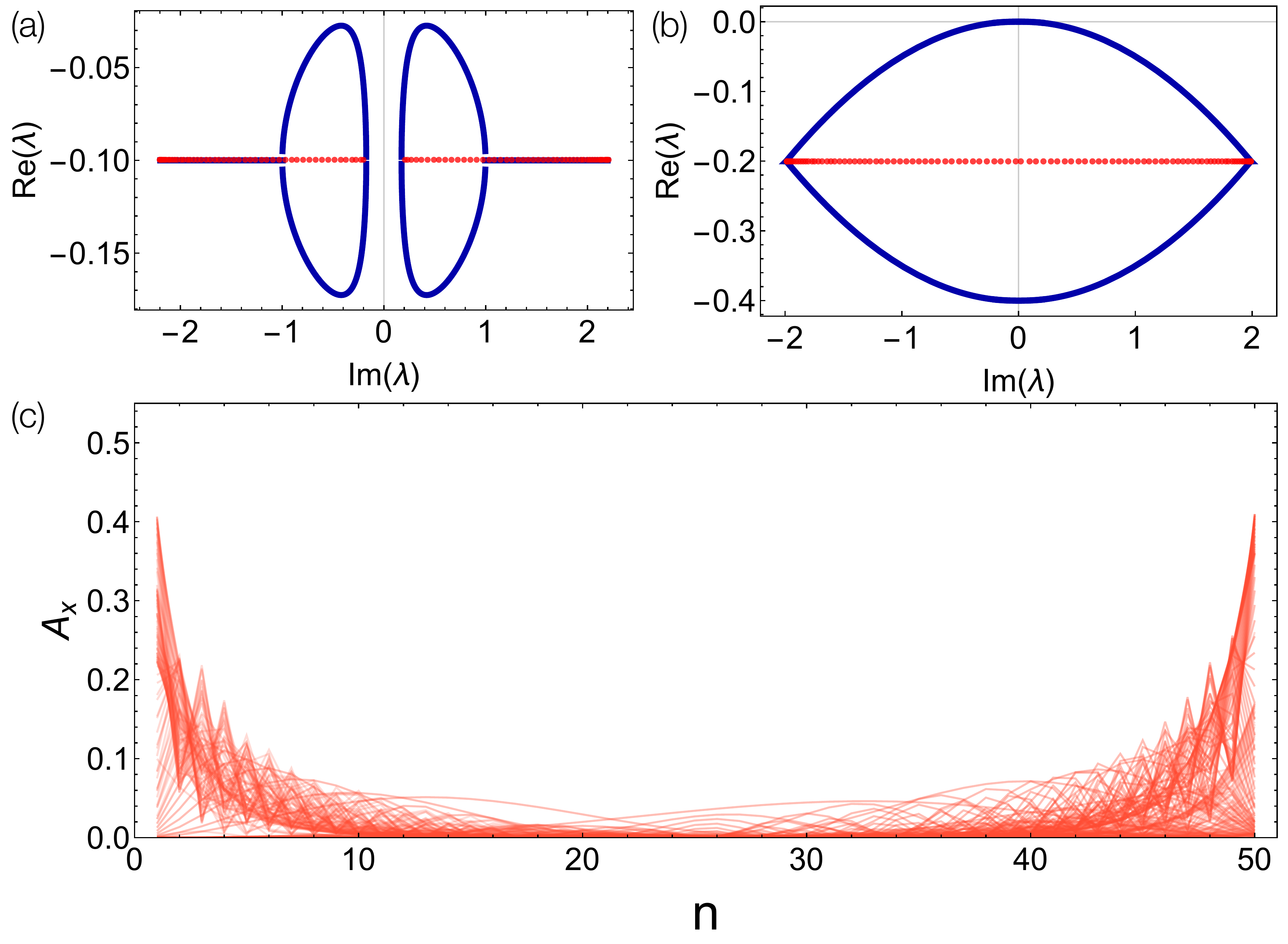}
\caption{(a), (b) Red (Blue) dots denote open (periodic) boundary spectrum of $X$. The parameter values are $t_1=1.2$, $t_2=1$, $\gamma=0.2$, $\delta_1=0.1$, $\delta_2=0$ for (a) and $t_1=t_2=1, \gamma=0.4, \delta_1=0.1, \delta_2=0$ for (b). (c) Sum of modular squares of the wave function in each unit cell $A_x$ for the open boundary damping matrix with the same parameters as in (b).}
\label{fig1}
\end{figure}

    Define $\Delta_{m,n}=Tr(c^{\dagger}_{m}c_{n}\rho)$ with $m,n=(x,s,o)$, and $\tilde \Delta=\Delta-\Delta_s$ with $\Delta_s$ denoting the stead value of $\Delta$.
   After some derivations \cite{SM}, the dynamical evolution of $\tilde \Delta$ is governed by
    \begin{equation}
     \frac{d \tilde \Delta(t) }{dt} = X \tilde \Delta (t) + \tilde \Delta(t) X^{\dagger},
    \end{equation}
    which gives rise to  $\tilde \Delta(t)=e^{Xt}\tilde \Delta(0)e^{X^{\dagger}t}$ with the damping matrix in the momentum space given by
     \begin{eqnarray}
        X &=& i\left[ \begin{array}{cc}
            H_{nSSH}(k) +\frac{i\gamma}{2}& \delta_1 \sin k +\delta_2 \sigma_z\\
            \delta_1 \sin k+\delta_2 \sigma_z& H_{nSSH}^T(-k)+\frac{i\gamma}{2}
            \end{array}
            \right ] \nonumber \\
             &=&(-\frac{\gamma}{2}+it_1\sigma_x+\frac{\gamma}{2}\sigma_y\tau_z)+i(t_2\sigma_y+\delta_1\tau_x) \sin k \nonumber\\
             & &+it_2\sigma_x \cos k++\delta_2\sigma_z\tau_x.  \label{xmatrix}
        \end{eqnarray}
      where  $\Delta_s=\frac{\gamma_g}{\gamma}\mathbb{I}$ with $\gamma=\gamma_l+\gamma_g$ and
   \begin{equation}
   H_{nSSH}(k)=(t_1+t_2 \cos k)\sigma_x+(t_2 \sin k-\frac{i\gamma}{2})\sigma_y
   \end{equation}
takes the same form of the non-Hermitian Su-Schrieffer-Heeger (SSH) model \cite{TELee,Yin,SYao1}.

When $\delta_2=0$, $X$ has anomalous time reversal symmetry, as it fulfills $CX(-k)^T=X(k)C$ with $C=i\tau_y$ \cite{CHLiu2}. We can get the eigenvalues of $X$  under open  boundary condition (OBC) and periodic boundary condition (PBC) as shown in Fig.\ref{fig1}(a) and (b). The mismatching of eigenvalues under open and periodic boundary is a characteristic sign of skin effect.
Define the sum of modular squares of X's eigen-wavefunction in each unit cell as $A_x=\sum_{o,s}|(\Psi_{x,o}^s)|^2$. In Fig.\ref{fig1}(c), we show the distribution of $A_x$ under the OBC. All the eigenstates of $X$ are localized on left and right boundaries, which is another sign of skin effect. If we put two identical models together and add a small symmetry-allowed perturbation, the skin effect is disappeared (SII in \cite{SM}). This is the characteristic of $\mathbb{Z}_2$ skin effect.

\begin{figure}[t]
    \begin{centering}
    \includegraphics[width=1\linewidth]{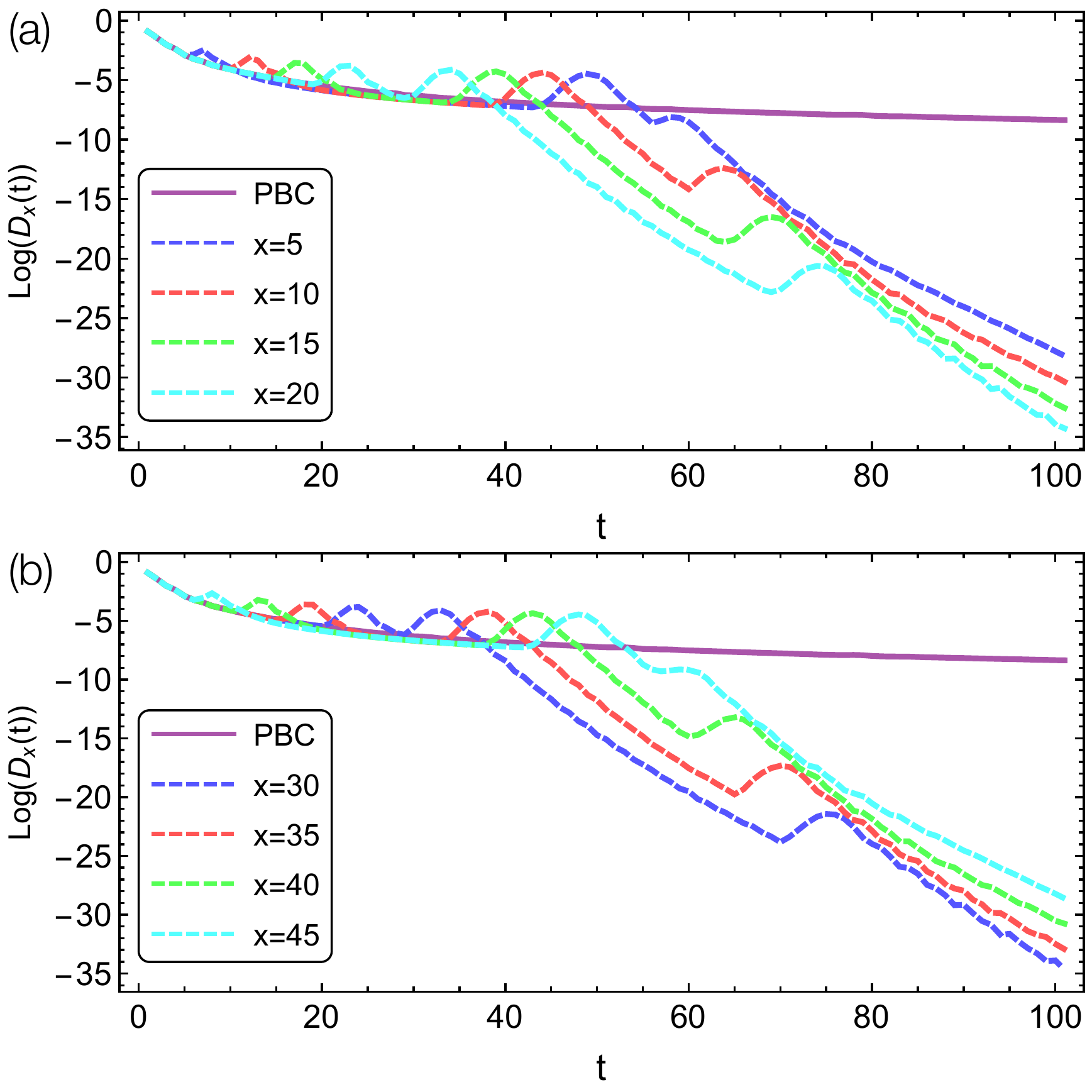}
    \par\end{centering}
    \caption{\label{fig2}(a) and (b) illustrate site-resolved relative local particle number damping of the open boundary chain and periodic chain (solid curve). The parameters are set as $t_1=t_2=1$, $\gamma=0.4$, $\delta_1=0.1$ and $\delta_2=0$.}
\end{figure}

\begin{figure}[t]
    \includegraphics[width=0.98\linewidth]{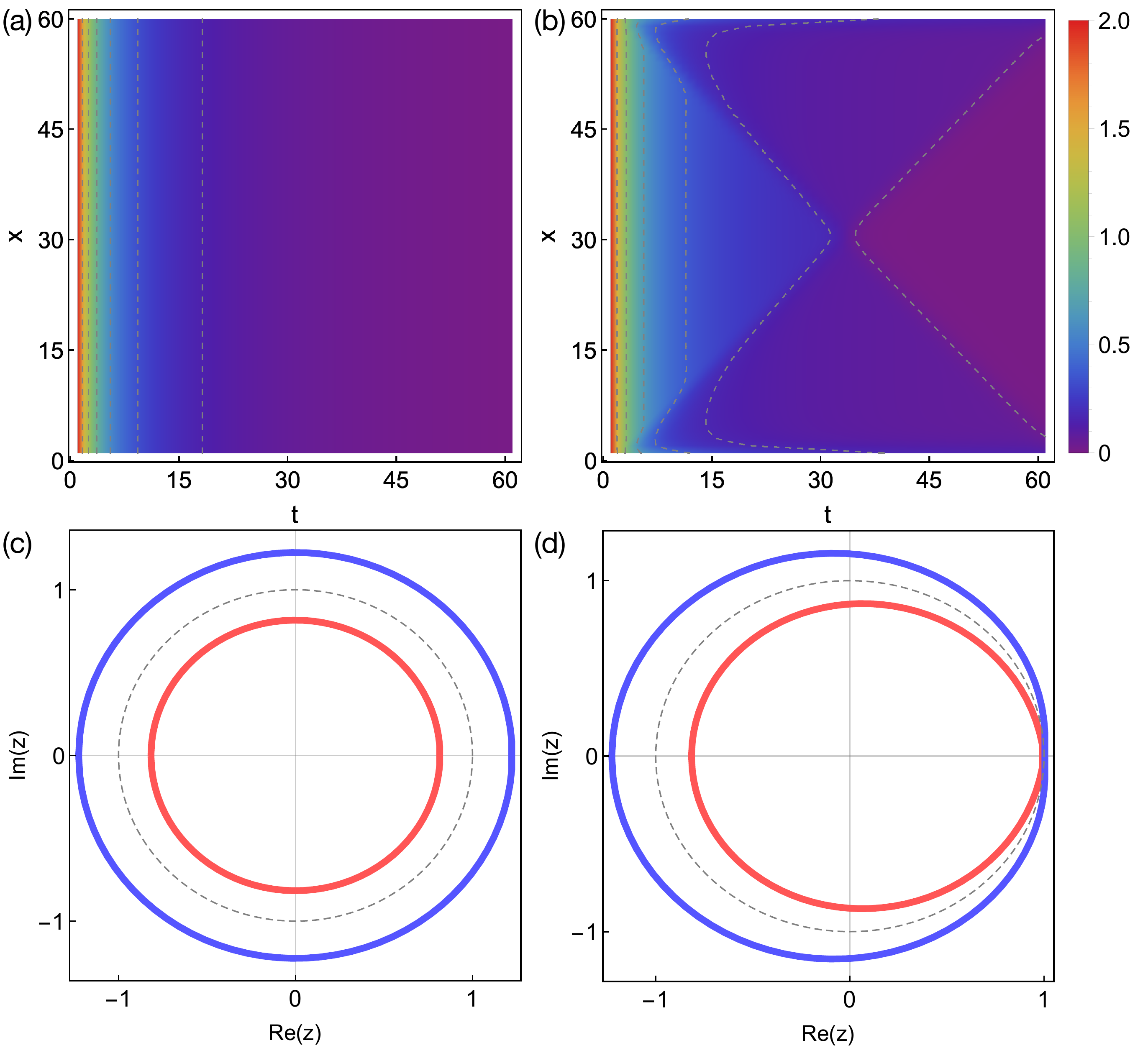}
    \caption{(a) and (b) display the time evolution of $ \tilde n_x(t)$ for the periodic and open boundary chain, respectively. (c) and (d) show the GBZ of damping matrix $X$, whereas the solid line shows GBZ and the dotted line shows the BZ, which is a unit circle given by the trajectory of  $|z|=1$.  The parameters are $t_1=t_2=1$, $\gamma=0.4$, $\delta_1=0.1$, $\delta_2=0$ for (a)(b)(d), and $t_1=t_2=1$, $\gamma=0.4$, $\delta_1=\delta_2=0$ for (c).
    }
    \label{fig3}
\end{figure}

 Given $n_{x,s,o} \equiv \Delta_{(x,s,o),(x,s,o)}$ denoting the particle number with spin $s$ and orbit $o$ at site $x$, we define the local damping as $D_{x}(t)=\sqrt{\sum_{s,o}(\frac{dn_{x,s,o}(t)}{dt})^2}$ and relative local particle number $\tilde n_x(t)=\sum_{s,o}\tilde\Delta_{(x,s,o),(x,s,o)}$. In Fig.\ref{fig2} (a) and (b) we display $log(D_x(t))$ as a function of $t$ for different $x$. While $D_{x}(t)$ under PBC is always a power law function of $t$, $D_{x}(t)$ under OBC changes from a power law function to an exponential function of $t$ during the evolution. We find that the transition time $t_c$ decreases as x increases for $0<x<20$, and
increases as x increases for $30<x<50$. In order to see more clearly the dependencies between $t_c$ and $x$, we plot the relative local particle number evolution in Fig.\ref{fig3}(a) and (b) for the periodic and open boundary system, respectively. We find that there are three main colors: dark blue, blue and purple, which are separated by two straight lines as shown in Fig.\ref{fig3}(b). The separatrix of the dark blue area and the purple area is the transition line. Such a phenomena is dubbed as helical damping. Nevertheless, the $\mathbb{Z}_2$ skin effect is not the sufficient condition of helical damping (see SIII in \cite{SM}), and we also require the Liouvillian gap of periodic lattice to be zero and the open boundary Liouvillian gap to be nonzero, where the Liouvillian gap is defined as $\Lambda_g=min[2 Re(-\lambda_n)]$ with $\lambda_n$ the eigenvalues of $X$. We notice that $\frac{[\tilde n_x(t)]_{OBC}}{[\tilde n_x(t)]_{PBC}}$ may exist helical behavior even if periodic boundary system is gapped\cite{SM}.
When the periodic boundary system is gapped (gapless), the short-time behavior of damping fulfills exponential (power) law for both the periodic and open boundary systems, since it costs time for sites located not on the boundary to get the boundary information. On the other hand, long-time behavior of OBC's damping fulfills exponential (power) law when the open boundary system is gapped (gapless), which will be explained further below.

Now we use non-Bloch band theory to explain helical damping.  For open boundary system the bulk wave function and eigenvalue of $X$ matrix can be obtained by replacing $X(k)$ with $X(k+i\kappa)$. All possible values of $e^{i(k +i\kappa)}$ constitute GBZ. In Fig.\ref{fig3}(c) and (d), we display the GBZ of the system with different parameters, which is composed of two closed curves with one inside and one outside the Brillouin zone (BZ). The relative local particle number can be decomposed into every GBZ modes: $\tilde n_{k+i\kappa}=\sum_{s,o} \langle x,s,o|e^{X(k+i\kappa)t}e^{X^{\dagger}(k+i\kappa)t}|x,s,o \rangle$. It follows
$max \left\{\tilde n_{k+i\kappa}\right\}\propto e^{2\kappa(x_1-x)-\gamma t}$ \cite{SM}. Define the velocity of open boundary system as $v_{k+i\kappa,\alpha}=Re(\frac{i\partial \lambda_{k+i\kappa,\alpha}}{\partial k})$, where $\alpha$ is band index, $\lambda_{k+i\kappa,\alpha}$ is the eigenvalue of $X(k+i\kappa)$ corresponding to $\alpha$ band. For simplicity we use $v$ to label $v_{k+i\kappa}$ in the following text. If the parameter settings are the same as in Fig.\ref{fig3}(b), we can get $v_{max}=-v_{min}\approx 1$ and $\kappa_{max}=-\kappa_{min}= 0.2$ at $k=\pi$ \cite{SM}. For $v=1$, $\kappa=-0.2$ and $x_1-x=-vt$, $x=x_1+vt\ge t$ and $e^{2\kappa(x_1-x)-\gamma t}=e^{(-2v\kappa-\gamma)t}=1$.  The decay factor cancels out, particle number damping fulfills a power law. Similarly, for $v=-1$, $\kappa=0.2$ and $x=x_1+vt\le L-t$, the decay factor also cancels out. For $x=x_1+vt\textless t$ and $x=x_1+vt\textgreater L-t$, this factor cannot be canceled out, and relative particle number damping obeys an exponential law. Due to the anomalous time reversal symmetry, we have $n_{x}(t)=n_{L-x}(t)$, which distributes symmetrically about $x=\frac{L}{2}$.

\begin{figure}[t]
\includegraphics[width=0.7\linewidth]{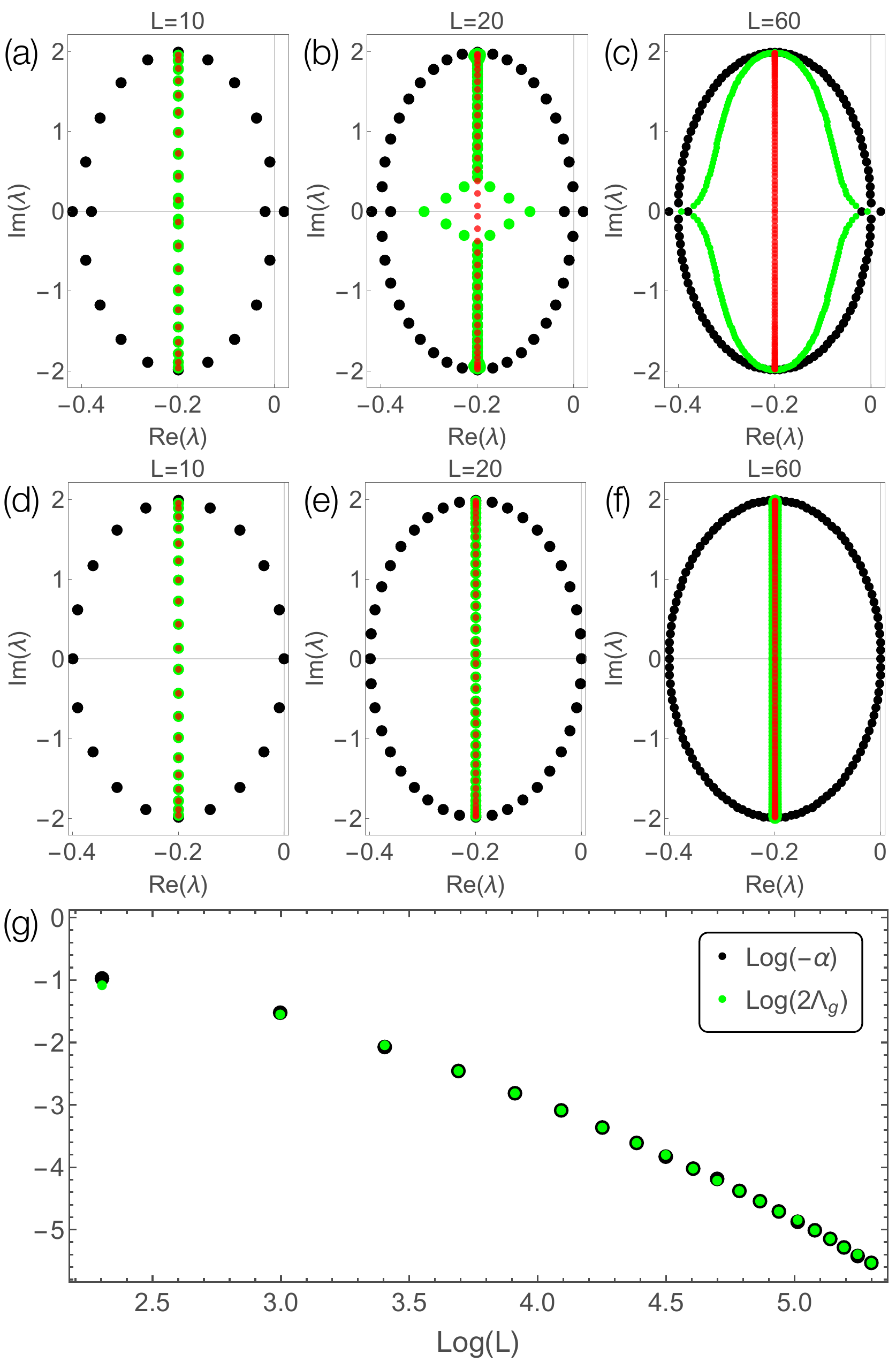}
\caption{\label{cse1} Open (green) and periodic (black) boundary spectrum of X for systems with  $ t_1= t_2 = 1, \gamma= 0.4$ and other parameters set as: $\delta_1=0, \delta_2 = 0.02$ for (a), (b) and (c) with different sizes, and  $\delta_1 = 0.02, \delta_2 = 0$ for (d), (e) and (f) with different sizes, respectively. The red dots denote the open boundary spectrum of X with parameters set as $t_1 = t_2 = 1, \gamma= 0.4, \delta_1= \delta_2 = 0$. (g) Long-time damping slopes $\alpha$ and $-2 \Lambda_g$ as a function of size $L$ with parameters set as $t_1 = t_2 = 1$, $\gamma= 0.4$, $\delta_1 = 0$, $\delta_2 = 0.02$.}
\end{figure}

\emph{Dynamic Critical Skin Effect.}--- When the system exhibits CSE, the open boundary energy spectrum is not continuous under the small change of parameters in the thermodynamic limit. For the finite size system, the open boundary spectrum is always continuous under the small change of parameters. Therefore, if CSE occurs, the energy spectrum of the system varies greatly with the size of the system. Here we study whether the perturbation $\delta_1$ or $\delta_2$ will cause CSE. And we want to detect this effect in dynamic experiments. With parameter set as $\delta_1=0.02$, $\delta_2=0$ or $\delta_1=0$, $\delta_2=0.02$, we calculate the spectrum of $X$ of the system with different sizes. The result is shown in Fig.\ref{cse1}. In Fig.\ref{cse1}(a)-(c), we set $\delta_1=0$, $\delta_2=0.02$ and display the spectrum of damping matrix for different system sizes. While the periodic spectrum is not sensitive to the system size $L$, the obvious change of open boundary spectrum with the increase of $L$  indicates the existence of CSE,  and the open boundary Liouvillian gap $\Lambda_g$ decreases as the system size increases. 

To measure the Liouvillian gap from dynamic experiment, we will deduce the relationship between relative particle number evolution and Liouvillian gap. The relative particle number is $\tilde n_x(t)=\sum_{i,j,s,o}e^{(\lambda_i+\lambda_j^*)t} \langle x,s,o|\Psi_i \rangle_{LR} \langle  \Psi_i|\Psi_j\rangle_{LR} \langle \Psi_j|x,s,o \rangle$, where subscripts $R$ and $L$ denote the right and left eigenvectors of $X$. Consider the case with large enough $t$. In this case, modes with $-Re(\lambda_i+\lambda_j^*)>2\Lambda_g$ can be omitted, and it follows $\tilde n_x(t)\approx ce^{-2\Lambda_gt}$. Assume that $log(\tilde n_x(t)) = \alpha(t)t+\beta$, then $ \alpha \approx -2\Lambda_g$. We numerically calculate the values of $2\Lambda_g$ and $\alpha$ for different size systems. As illustrated in Fig.\ref{cse1}(g), the numerical results are consistent with our theoretical analysis. We also analyze the scaling of the Liouvillian gap with the system size, which indicates $log(2\Lambda_g)\approx  -2.3log(L)+6.8$ around $L = 200$ and the absolute value of this slope increases as $L$ increases. When the system is large enough that $-2\Lambda_g >\alpha_{max}$, the helical damping is hidden. When the system is small enough that $-2\Lambda_g < \alpha_{min}$, the helical damping is manifested. Here $\alpha_{max}/\alpha_{min}$ is the maximum/minimum slope of $log(\tilde n_x(t))$ in the power law interval. In Fig.\ref{cse1}(d)-(f), we set $\delta_1= 0.02, \delta_2 = 0$. It is clearly shown that there is no CSE, and the open boundary Liouvillian gap $\Lambda_g(L)$ does not change as the system size increases. Therefore we can detect the presence of CSE by measuring the damping spectra of systems of different sizes in this case. Remarkably, we construct an example that two irreducible subsystems with different GBZs are coupled together but no CSE occurs. Specifically, $X$ is constructed by coupling two systems $iH_{nSSH}(k)-\frac{\gamma}{2}$ and $iH^T_{nSSH}(-k)-\frac{\gamma}{2}$, which have different GBZs. The perturbation term of $\delta_1$ couples $iH_{nSSH}(k)-\frac{\gamma}{2}$ and $iH^T_{nSSH}(-k)-\frac{\gamma}{2}$, but there is no CSE.

Here we give an explanation in terms of GBZ. The bulk spectrum of open boundary $X$ is composed of eigenvalues of $X(k+i\kappa)$, and thus it is a continuous function of GBZ. On the other hand, the solution of a certain GBZ equation is a continuous function of parameters of $X$. Therefore the only origin of energy spectrum's discontinuity is the change of GBZ equation. For case 1 with $\delta_1=\delta_2=0$,  $X$ is reducible, and the characteristic polynomial is $f_1(z,\lambda)=det(iH_{nSSH}-\frac{\gamma}{2}-\lambda I)$ and $f_2(z,\lambda)=det(iH_{nSSH}^T-\frac{\gamma}{2}-\lambda I)$, where $z = e^{ik}$. The roots of $f_1=0$ and $f_2=0$ are $z^a_1, z^a_2$ and $z^b_1, z^b_2$, respectively, where $|z^a_1|\le |z^a_2|$, $z^b_i={1}/{z^a_i}$ ($i=1,2$). The GBZ equations are $|z^a_1|=|z^a_2|$ and $|z^b_1|=|z^b_2|$ \cite{KYokomizo,KZhang}. If $X$ is irreducible non-Hermitian matrix, the characteristic polynomial is $f(z,\lambda) = det(X- \lambda I)$, and the solution of $f(z,\lambda) = 0$ is $z_1,z_2,z_3,z_{4}$, where $|z_1|\le |z_2| \le |z_3|\le |z_{4}|$. For case 2 with $\delta_1=0$ and $\delta_2\ne 0$, $X$ does not have any symmetry, and the GBZ equation is $|z_{2}| =| z_{3}|$. For case 3 with $\delta_1\ne 0$ and $\delta_2=0$, X has anomalous time reversal symmetry. The GBZ equations are $|z_{1}| = |z_{2}|$ and $|z_{3}| = |z_{4}|$ and the roots satisfy $z_{2+i}={1}/{z_{3-i}}$ ($i=1,2$) \cite{YYi}. We emphasize that the GBZ equation of case 1 and case 3 are equivalent \cite{SM}. So there is a change of GBZ equation from case 1 to case 2, which causes the discontinuity of eigenvalues and wave functions under the time-reversal-breaking perturbation. The GBZ equation will not change from case 1 to case 3, therefore CSE does not occur in this process.

\emph{Anomalous Critical Skin Effect.}---Here we construct a model that couples two irreducible subsystems with same GBZs but shows CSE. It is anomalous critical skin effect. We consider the model described by
\begin{equation}
\begin{split}
X_1=& \left[ \begin{array}{cc}
X & i\delta_3 \sigma_x\\
i\delta_3 \sigma_x& X
\end{array}
\right ] \\ \label{x1matrix}
\end{split} .
\end{equation}
The parameters are set as $t_1 = t_2 = 1, \gamma= 0.4, \delta_1 = 0.1, \delta_2 = 0$ and $\delta_3 = 0.01$. It couples two identical systems of X described by Eq.(\ref{xmatrix}), but exhibits CSE (See SII in \cite{SM}).

When $\delta_3=0$, the GBZ equations of $X_1$ are $|z_{1}|=|z_{2}|$ and $|z_{3}|=|z_{4}|$. When $\delta_3\ne 0$, the characteristic polynomial of $X_1$ is $g(z)=det(X_1( z)-\lambda I)$, with the solution of $g( z,\lambda) = 0$ given by $\tilde z_1,\tilde z_2,...,\tilde z_{8}$, where $|\tilde z_1|\le |\tilde z_2| \le ... \le |\tilde z_{8}|$. The GBZ equations are $|\tilde z_{3}|=|\tilde z_{4}|$ and $|\tilde z_{5}|=|\tilde z_{6}|$. The GBZ equation also changes when CSE occurs. The reason for the discontinuity is the change in GBZ equation. Furthermore, changes in GBZ equation require changes of symmetries of the system or  number of roots of characteristic equation. In this case, when $\delta_3$ changes from zero to non-zero, the unitary symmetry $\sigma_z X_1=X_1\sigma_z$ disappears. Therefore, changes in GBZ equation are necessary conditions for CSE, but the reverse is not correct. An example is that no matter how GBZ equation changes for a Hermitian system, there will be no CSE. The symmetry $H^{\dagger}=H$ keeps the GBZ to be still a unit circle, even if the GBZ equation changes.

\emph{Conclusion.}---When the open boundary spectrum of the damping matrix of system is gapped and the periodic boundary spectrum is gapless, this system has a non-trivial damping spectrum. If the system also has $\mathbb{Z}_2$ skin effect, helical damping occurs. Adding perturbations which break anomalous time reversal symmetry, CSE occurs and helical damping disappears for systems in the thermodynamic limit, but still exists for small size systems. Coupling two identical models with $\mathbb{Z}_2$ skin effects by perturbation, we can realize anomalous critical skin effect. And we unveil the origin of the discontinuity coming from the change of GBZ equation. These phenomena can be verified in open quantum system by measuring local particle number evolution.

\begin{acknowledgments}
The work is supported by NSFC under Grants No.11974413, the National Key
Research and Development Program of China (2016YFA0300600 and
2016YFA0302104) and the Strategic Priority Research Program of Chinese Academy of Sciences under Grant No. XDB33000000.
\end{acknowledgments}

\newpage

\onecolumngrid
\begin{center}
	{\bf \large Supplemental Material for \\ \smallskip `` Helical damping and anomalous critical non-Hermitian skin effect "}
\end{center}

\section{SI: Derivation of evolution equation of single-particle correlation matrix}
Next we derive the expression of  evolution equation of single-particle correlation matrix $\Delta$ with $\Delta_{mn}=Tr(c_{m}^{\dagger} c_{n}\rho)$. Inserting the Lindbald equation into into
\[
\left(\frac{d\Delta_{mn}}{dt}\right)= Tr(c_{m}^{\dagger} c_{n}\frac{d\rho}{dt}) ,
\]
we have
\begin{equation}
\begin{split}
\left(\frac{d\Delta_{mn}}{dt}\right)
=&Tr\left[c_{m}^{\dagger} c_{n}\left(-i[H,\rho]+\sum _{\mu}(2L_{\mu}\rho L_{\mu}^{\dagger}-\left\{L_{\mu}^{\dagger}L_{\mu},\rho\right\})\right)\right] \\
=&Tr\left[-ic_{m}^{\dagger} c_{n}H\rho+ic_{m}^{\dagger} c_{n}\rho H+\sum_{\mu}\left(2c_{m}^{\dagger} c_{n}L_{\mu}\rho L_{\mu}^{\dagger}-c_{m}^{\dagger} c_{n}L_{\mu}^{\dagger}L_{\mu}\rho-c_{m}^{\dagger} c_{n}\rho L_{\mu}^{\dagger}L_{\mu}\right)\right] \\
=&Tr\left\{\left[-ic_m^{\dagger} c_{n}H+iHc_{m}^{\dagger} c_{n} +\sum_{\mu}\left(2L_{\mu}^{\dagger}c_{m}^{\dagger} c_{n}L_{\mu} -c_{m}^{\dagger} c_{n}L_{\mu}^{\dagger}L_{\mu}-L_{\mu}^{\dagger}L_{\mu}c_{m}^{\dagger} c_{n}\right)\right]\rho\right\}  \\
=&Tr\left\{\left[-i[c_{m}^{\dagger}c_{n},H]+\sum_{\mu}\left(2L_{\mu}^{\dagger}c_{m}^{\dagger} c_{n}L_{\mu}-2L_{\mu}^{\dagger}L_{\mu}c_{m}^{\dagger} c_{n} +L_{\mu}^{\dagger}L_{\mu}c_{m}^{\dagger} c_{n}-c_{m}^{\dagger} c_{n}L_{\mu}^{\dagger}L_{\mu}\right)\right]\rho\right\} \\
=&Tr\left\{\left[-i[c_{m}^{\dagger}c_{n},H]+\sum_{\mu}\left(2L_{\mu}^{\dagger}[c_{m}^{\dagger} c_{n},L_{\mu}]+[L_{\mu}^{\dagger}L_{\mu},c_{m}^{\dagger} c_{n}]\right)\right]\rho\right\} .
\end{split} \label{delta}
\end{equation}
Here $H=\sum_{j,k}h_{jk}c_{j}^{\dagger}c_{k}$, $L_{\mu}=L^g_{\mu}$ or $L^l_{\mu}$ with $L_\mu^g=\sum_{k}D_{\mu k}^gc^{\dagger}_{k}$ and $L_\mu^l=\sum_{k}D_{\mu k}^lc_{k}$, $j,k,m,n$, is fermion index and $\mu,\nu$ is Lindblad operator's index. And we define $M_{jk}^g=\sum_{\mu}D_{\mu j}^{g*}D_{\mu k}^g$ and $M_{jk}^l=\sum_{\mu}D_{\mu j}^{l*}D_{\mu k}^l$. The first term gives:
\begin{equation}
\begin{split}
Tr(-i[c_{m}^{\dagger}c_{n},H]\rho)=&\sum_{j,k}Tr(-ih_{jk}[c_{m}^{\dagger}c_{n},
c_{j}^{\dagger}c_{k}]\rho)\\
=&\sum_{j,k}-ih_{jk}Tr(c_{m}^{\dagger}\left\{c_{n},
c_{j}^{\dagger}\right\}c_{k}\rho-
c_{j}^{\dagger}\left\{c_{k},c_{m}^{\dagger}\right\}c_{n}\rho) \\
=&\sum_{j,k}-ih_{jk}Tr(\delta_{n,j}c_{m}^{\dagger}c_{k}\rho-
\delta_{k,m}c_{j}^{\dagger}c_{n}\rho) \\
=&\sum_{k}-ih_{nk}Tr(c_{m}^{\dagger}c_{k}\rho)+
\sum_{j}ih_{jm}Tr(c_{j}^{\dagger}c_{n}\rho) \\
=&\sum_{k}(-ih_{nk}\Delta_{mk}+ih_{km}\Delta_{kn}), 
\end{split} \label{term1}
\end{equation}
the second term gives
\begin{equation}
\begin{split}
\sum_{\mu}Tr(2L_{\mu}^{\dagger}[c_{m}^{\dagger} c_{n},L_{\mu}]\rho)=&
\sum_{\mu}Tr(2L_{\mu}^{g\dagger}[c_{m}^{\dagger} c_{n},L^g_{\mu}]\rho+2L_{\mu}^{l\dagger}[c_{m}^{\dagger} c_{n},L^l_{\mu}]\rho)  \\
=&\sum_{\mu jk}Tr(2D_{\mu j}^{g*}D^g_{\mu k}c_{j}[c_{m}^{\dagger} c_{n},c_{k}^{\dagger}]\rho+
2D_{\mu j}^{l*}D^l_{\mu k}c_{j}^{\dagger}[c_{m}^{\dagger} c_{n},c_{k}]\rho)  \\
=&\sum_{\mu jk}Tr(2D_{\mu j}^{g*}D^g_{\mu k}c_{j}c_{m}^{\dagger} \left\{ c_{n},c_{k}^{\dagger}\right\}\rho-
2D_{\mu j}^{l*}D^l_{\mu k}c_{j}^{\dagger}\left\{c_{k},c_{m}^{\dagger}\right\} c_{n}\rho)  \\
=&\sum_{\mu jk}Tr(2\delta_{n,k}D_{\mu j}^{g*}D^g_{\mu k}c_{j}c_{m}^{\dagger} \rho-
2\delta_{k,m}D_{\mu j}^{l*}D^l_{\mu k}c_{j}^{\dagger} c_{n}\rho)  \\
=&\sum_{\mu j}Tr(2D_{\mu j}^{g*}D^g_{\mu n}c_{j}c_{m}^{\dagger} \rho-
2D_{\mu j}^{l*}D^l_{\mu m}c_{j}^{\dagger} c_{n}\rho)  \\
=&\sum_{j}2M^g_{jn}(\delta_{mj}-\Delta_{mj})-
2M^l_{jm}\Delta_{jn} \\
=&2M^g_{mn}-\sum_{j}(2M^g_{jn}\Delta_{mj}+
2M^l_{jm}\Delta_{jn}),
\end{split} \label{term2}
\end{equation}
and the third term gives
\begin{equation}
\begin{split}
& \sum_{\mu}Tr\left([L_{\mu}^{\dagger}L_{\mu},c_{m}^{\dagger} c_{n}]\rho\right) \\
=
&\sum_{\mu jk}Tr\left(D_{\mu j}^{g*}D_{\mu k}^g[c_jc_k^{\dagger},c_{m}^{\dagger} c_{n}]\rho+
D_{\mu j}^{l*}D_{\mu k}^l[c_j^{\dagger}c_k,c_{m}^{\dagger} c_{n}]\rho\right)  \\
=&\sum_{\mu jk}Tr\left[D_{\mu j}^{g*}D_{\mu k}^g\left(-\left\{c_j,c_{m}^{\dagger}\right\}c_k^{\dagger}c_{n}+c_{m}^{\dagger}c_j\left\{c_k^{\dagger}, c_{n}\right\}\right)\rho+
D_{\mu j}^{l*}D_{\mu k}^l\left(c_j^{\dagger}\left\{c_k,c_{m}^{\dagger}\right\} c_{n}-c_{m}^{\dagger}\left\{c_{n},c_j^{\dagger}\right\}c_k\right)\rho\right]  \\
=&\sum_{jk}Tr\left[M_{jk}^g\left(-\delta_{j,m}c_k^{\dagger}c_{n}+\delta_{k,n}c_{m}^{\dagger}c_j\right)\rho+
M_{jk}^l\left(\delta_{k,m}c_j^{\dagger} c_{n}-\delta_{j,n}c_{m}^{\dagger}c_k\right)\rho\right]  \\
=&Tr\left[\sum_{k}-M_{mk}^gc_k^{\dagger}c_{n}\rho+\sum_{j}M_{jn}^lc_{m}^{\dagger}c_j\rho+
\sum_{j}M_{jm}^lc_j^{\dagger} c_{n}\rho-\sum_{k}M_{nk}^lc_{m}^{\dagger}c_k\rho\right]  \\
=&\sum_{k}(-M_{mk}^g\Delta_{kn}+M_{kn}^g\Delta_{mk}+
M_{km}^l\Delta_{kn}-M_{nk}^l\Delta_{mk}) .
\end{split} \label{term3}
\end{equation}
Combining them together, we get:
\begin{equation}
\begin{split}
\left(\frac{d\Delta_{mn}}{dt}\right)=&2M^g_{mn}+\sum_{k}(-ih_{nk}\Delta_{mk}+ih_{km}\Delta_{kn}-
2M^g_{kn}\Delta_{mk}-
2M^l_{km}\Delta_{kn}-M_{mk}^g\Delta_{kn}+M_{kn}^g\Delta_{mk}+
M_{km}^l\Delta_{kn}-M_{nk}^l\Delta_{mk})  \\
=&2M^g_{mn}+\sum_{k}(-ih_{nk}\Delta_{mk}+ih_{km}\Delta_{kn}-
M^g_{kn}\Delta_{mk}-
M^l_{km}\Delta_{kn}-M_{mk}^g\Delta_{kn}-M_{nk}^l\Delta_{mk})  \\
=&\left(2M^g+i\left[h^T,\Delta\right]-\left\{M^g+M^{lT},\Delta\right\}\right)_{mn} \\
=&\left(2M^g+X\Delta+\Delta X^{\dagger}\right)_{mn} ,
\end{split}
\end{equation}
where $X=ih^T-(M^g+M^{lT})$. For $h$ and $L_{\mu}$ given by Eq.(2) and (3) in the main text, we can calculate that $M^g=\frac{\gamma_g}{2}(\sigma_0\tau_0-\sigma_y\tau_z)$, $M^l=\frac{\gamma_l}{2}(\sigma_0\tau_0+\sigma_y\tau_z)$ and
$X=it_1\sigma_x+i(t_2\sigma_y+\delta_1\tau_x) \sin k+i(t_2\sigma_x+\delta_2\sigma_y\tau_x) \cos k
-\frac{\gamma_l+\gamma_g}{2}(\sigma_0\tau_0-\sigma_y\tau_z)$, which gives  Eq.(5) in the main text.
Let $\frac{d\Delta}{dt}=2M^g+X\Delta+\Delta X^{\dagger}=0$,
the solution is $\Delta_s=\frac{\gamma_g}{\gamma}\mathbb{I}$.
We can verify that $2M^g+X\Delta_s+\Delta_s X^{\dagger}=\gamma_g(\sigma_0\tau_0-\sigma_y\tau_z)-2\frac{\gamma_g}{\gamma}\frac{\gamma_l+\gamma_g}{2}(\sigma_0\tau_0-\sigma_y\tau_z)=0$

\section{SII: Model for exhibiting anomalous critical skin effect}
\begin{figure}[h]
\includegraphics[width=5in]{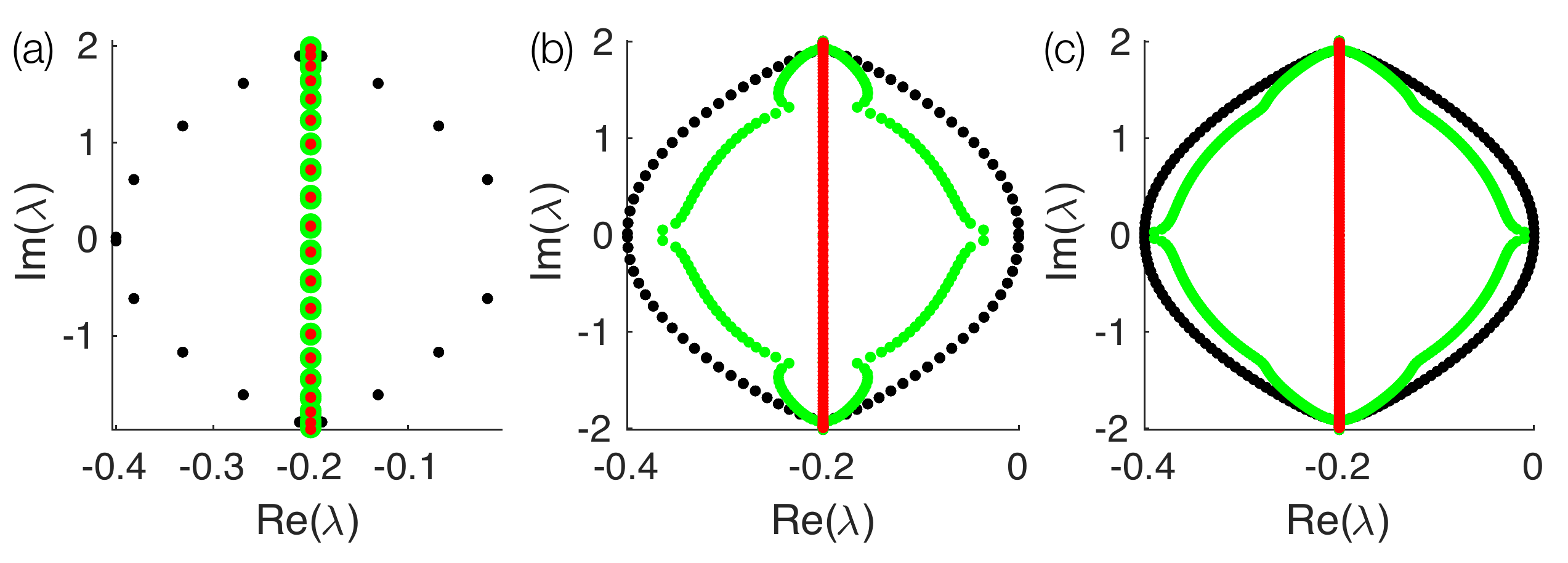}\\
\caption{ Open (green) and periodic (black) boundary spectrum with parameters $t_1 = t_2 = 1, \gamma= 0.4, \delta_1 = 0.1, \delta_2 = 0, \delta_3 = 0.01$ for (a), (b) and (c) with different sizes. The red dots denote the open boundary spectrum of $X_1$ with parameters $t_1 = t_2 = 1, \gamma= 0.4, \delta_1 = 0.1, \delta_2 = \delta_3 = 0$.
} \label{figsm1}
\end{figure}

In this section, we show that the skin effect and helical damping is not stable to symmetry-allowed perturbation $\delta_3$ in Eq.(7) of the main text. The considered damping matrix is given by
\begin{equation}
\begin{split}
X_1=& \left[ \begin{array}{cc}
X & i\delta_3 \sigma_x\\
i\delta_3 \sigma_x& X
\end{array}
\right ] . \\ \label{eqx1}
\end{split}
\end{equation}
The open boundary and periodic boundary spectrum of $X_1$ are displayed in Fig.\ref{figsm1}. While the periodic boundary spectrum is not sensitive to the lattice size, the shape of open boundary spectrum changes obviously with the increase in the lattice size. Such an obvious change of open boundary spectrum is induced by the perturbation term of $\delta_3$.

\begin{figure}[h]
\includegraphics[width=3in]{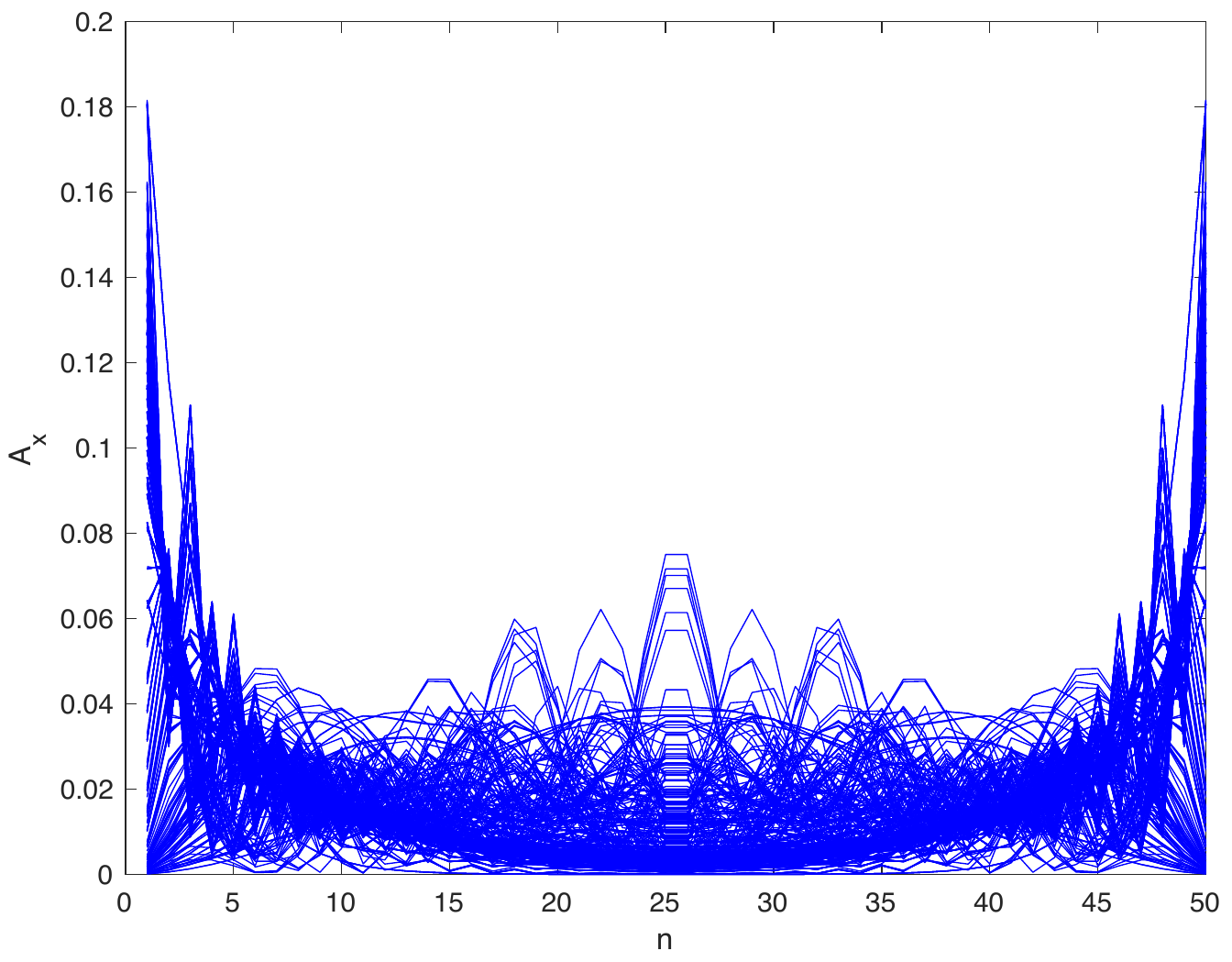}\\
\caption{  Sum of modular squares of the wave function in each unit cell $A_x$ for the open boundary damping matrix of $X_1$. The parameters of $X_1$ are set as: $t_1 = t_2 = 1, \gamma= 0.4, \delta_1 = 0.1,
\delta_2 = 0, \delta_3 = 0.01$. The system size is $L=50$. }\label{figsm2}
\end{figure}

In Fig.\ref{figsm2}, we show the distribution of $A_x$ under the OBC, where $A_x=\sum_{\alpha}|\Psi_{x\alpha}|^2$ is sum of modular squares of the amplitude of the wave function of of $X_1$ in each unit cell,  $\alpha$ is degrees of freedom in each cell and $x$ is the cell index.  It is clear that some eigenstates of $X_1$ spread over all the lattices. We find that there is no skin effect for this model.


\section{SIII: Example with $Z_2$ skin effect but no helical damping and generalized helical damping}
\begin{figure}[t]
\includegraphics[width=3in]{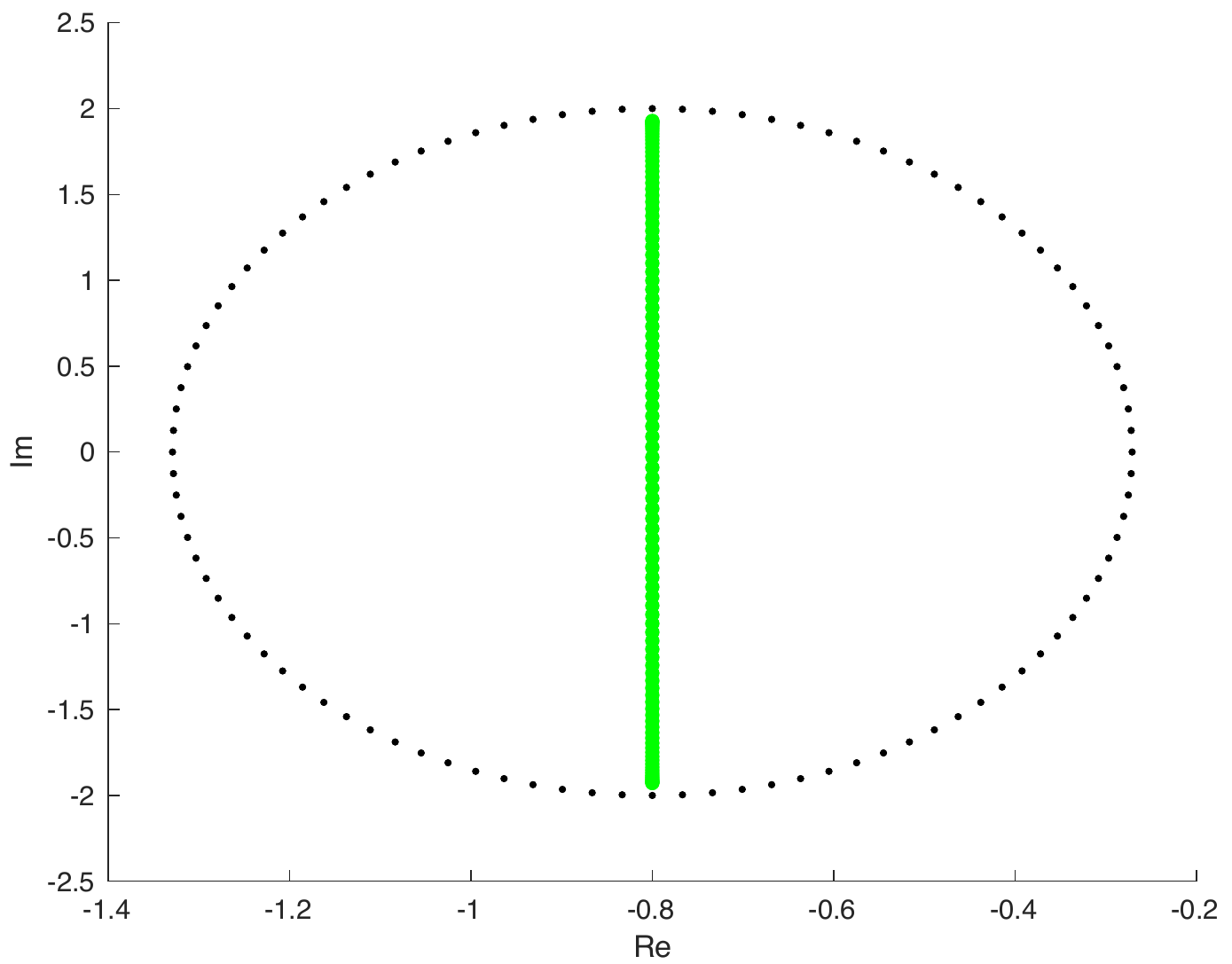}\\
\caption{Open (green) and periodic (black) boundary spectra of $X$ matrix. The parameters are taken as $t_1=1, \gamma_g=0.2, \delta=0.2$ with the system size L=100.} \label{figsm3}
\end{figure}

In this section,  we show that $Z_2$ skin effect is not a sufficient condition for the occurrence of helical damping of particle number. When particle number damping in the periodic boundary system fulfills an exponential law, the particle number damping in the open boundary system  always follows an exponential law. In this case,  the Liouvillian gap is not zero.
Consider a one-dimensional lattice with each cell having one orbit and spin degree of freedom. described by the following Hamiltonian in the momentum space:
\begin{equation}
h(k)=2t_1 \cos k_x +2\delta \tau_x ,
\end{equation}
where $\tau_{\alpha}$ ($\alpha=x,y,z$) act on spin space. Suppose that the Lindblad operators are given by
\begin{equation}
\begin{split}
L_{x1}^g=\sqrt{\frac{\gamma_g}{2}}(c^{\dagger}_{x\uparrow}-c^{\dagger}_{x\downarrow}+ic^{\dagger}_{x+1\uparrow}+ic^{\dagger}_{x+1\downarrow}), \\
L_{x2}^g=\sqrt{\frac{\gamma_g}{2}}(c^{\dagger}_{x\uparrow}+c^{\dagger}_{x\downarrow}+ic^{\dagger}_{x+1\uparrow}-ic^{\dagger}_{x+1\downarrow}),
\end{split}
\label{operators}
\end{equation}
it follows $\tilde \Delta(t) =\Delta(t)-\Delta_s=e^{Xt}\tilde \Delta(0)e^{X^{\dagger}t}$, where $X=2it_1 \cos k_x -2\gamma_g\tau_0+(2i\delta\tau_x+2\gamma_g\tau_z) \sin k$ and $\Delta_s=\mathbb{I}$. We display the spectrum of $X$ matrix under PBC and OBC in Fig.\ref{figsm3}, which indicates the existence of nonzero Liouvillian gap for both periodic and open boundary systems. The disappearance of skin modes, after putting two identical models together and adding a small symmetry-allowed perturbation, indicates the existence of $Z_2$ skin effect.
\begin{figure}[t]
\includegraphics[width=4in]{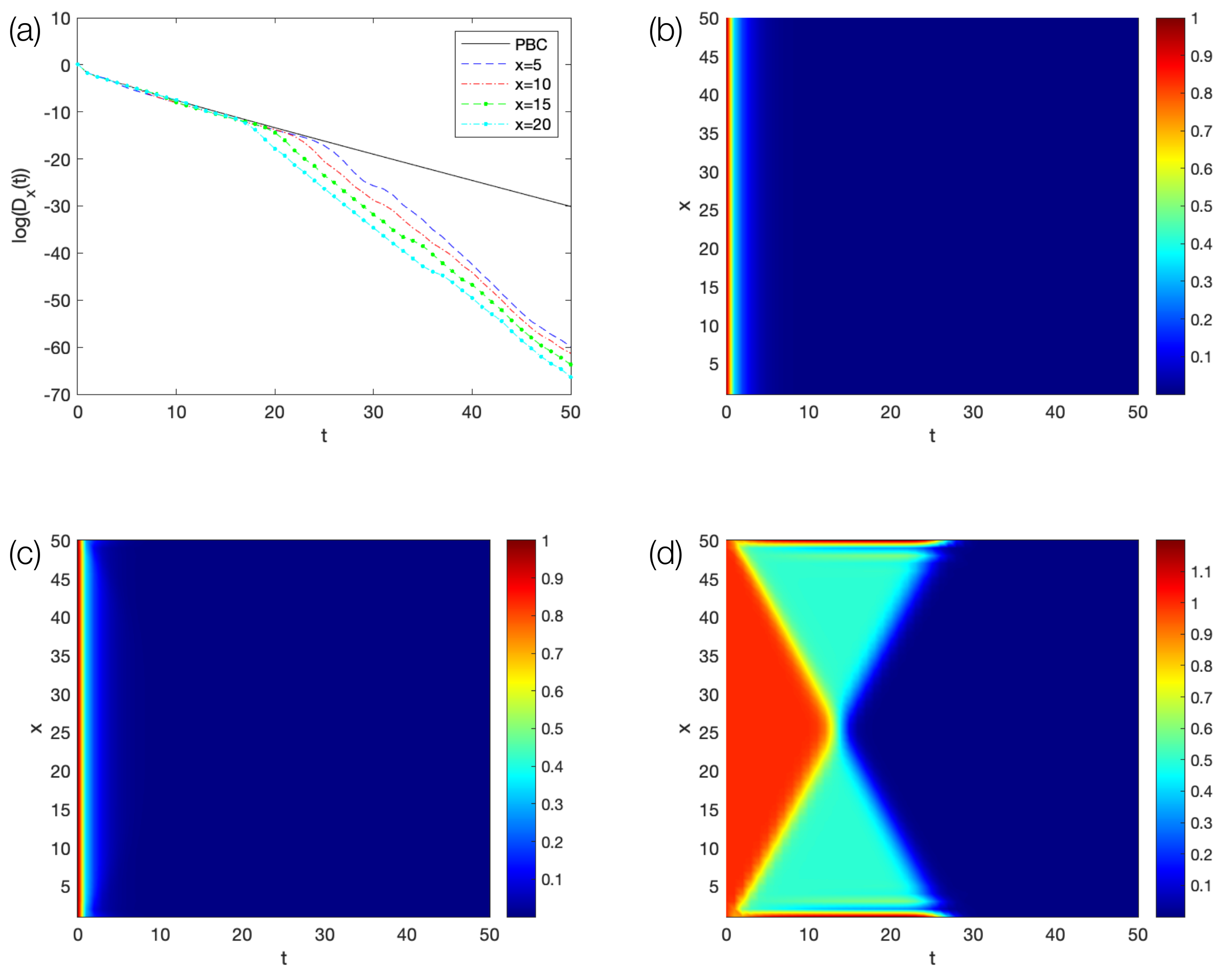}\\
\caption{(a) illustrates site-resolved relative local particle number damping for different cells. (b) and (c) show the evolution of $|\tilde n_x(t)|$ for system under (b) PBC and (c) OBC, respectively. (d) shows the evolution of $\frac{(|\tilde n_x(t)|)_{OBC}}{(|\tilde n_x(t)|)_{PBC}}$. The parameters are set as $t_1=1, \gamma_g=0.2, \delta=0.2$ with the system size $L=50$.} \label{figsm4}
\end{figure}

Set the initial state as the state without particle occupation, and we can get $\tilde \Delta=-\mathbb{I}$.
  We show the relative local particle number damping for different cells in Fig.\ref{figsm4}(a). And we calculate the evolution of $|\tilde n_x(t)|$ and display the numerical results in Fig.\ref{figsm4}(b) and (c), which indicates the particle number damping under both PBC and OBC fulfilling exponential law.  In Fig.\ref{figsm4}(d), we display the evolution of $\frac{(| \tilde n_x(t)|)_{OBC}}{(|\tilde n_x(t)|)_{PBC}}$, which exhibits helical behavior. And we call it { \it generalized helical damping}. The generalized helical damping is a more inclusive physical phenomenon than helical damping, and it don't need the periodic boundary Liouvillian gap to be gapless. The mismatch of open and periodic boundary damping spectrum is the necessary condition.

\section{SIV: The decay factor of relative particle number}
The $k+i\kappa$ component relative particle number in each cell is:
\begin{equation}
\begin{split}
\tilde n_{x,k+i\kappa}=&\sum_{s,o}\langle(x,s,o)|e^{X(k+i\kappa)t}e^{X^{\dagger}(k+i\kappa)t}|(x,s,o)\rangle\\
=&\sum_{s,o,x_1,s_1,o_1}\langle(x,s,o)|e^{X(k+i\kappa)t}|(x_1,s_1,o_1)\rangle \langle(x_1,s_1,o_1)|e^{X^{\dagger}(k+i\kappa)t}|(x,s,o)\rangle\\
=&\sum_{s,o,x_1,s_1,o_1}|\langle(x,s,o)|e^{X(k+i\kappa)t}|(x_1,s_1,o_1)\rangle|^2  \\
=&\sum_{\alpha}\sum_{s,o,x_1,s_1,o_1}|\langle(x,s,o)|e^{X(k+i\kappa)t}|k+i\kappa,\alpha\rangle_{RL}\langle k+i\kappa,\alpha|(x_1,s_1,o_1)\rangle|^2\\
=&\sum_{\alpha}\sum_{s,o,x_1,s_1,o_1}|\langle (s,o)|\alpha\rangle_{RL}\langle \alpha|(s_1,o_1)\rangle e^{i(k+i\kappa)(x_1-x)+\lambda_{k+i\kappa,\alpha}t}|^2\\
\propto &\sum_{\alpha}\sum_{s,o,x_1,s_1,o_1}|e^{i(k+i\kappa)(x-x_1)+\lambda_{k+i\kappa,\alpha}t}|^2 ,
\end{split}
\end{equation}
and
\begin{equation}
\begin{split}
 max \left\{\tilde n_{x,k+i\kappa}\right\}\propto & max \left\{\sum_{\alpha}\sum_{s,o,x_1,s_1,o_1}|e^{i(k+i\kappa)(x-x_1)+\lambda_{k+i\kappa,\alpha}t}|^2 \right\}\\
\propto &|e^{2i(k+i\kappa)(x-x_1)-\gamma t}|  \\
\propto &e^{2\kappa(x_1-x)-\gamma t} ,
\end{split}
\end{equation}
where $\alpha$ is band index, $\lambda_{k+i\kappa,\alpha}$ ($|k+i\kappa,\alpha\rangle_{R}$) is the eigenvalue (eigenvector) of $X(k+i\kappa)$ corresponding to $\alpha$ band, $max\left\{Re(\lambda_{k+i\kappa,\alpha})\right\}=-\frac{\gamma}{2}$, $\langle x|k+i\kappa \rangle_{R}=e^{i(k+i\kappa)x}$ and $_{L}\langle k+i\kappa|x_1 \rangle=e^{-i(k+i\kappa)x_1}$.

\begin{figure}[h]
\includegraphics[width=5in]{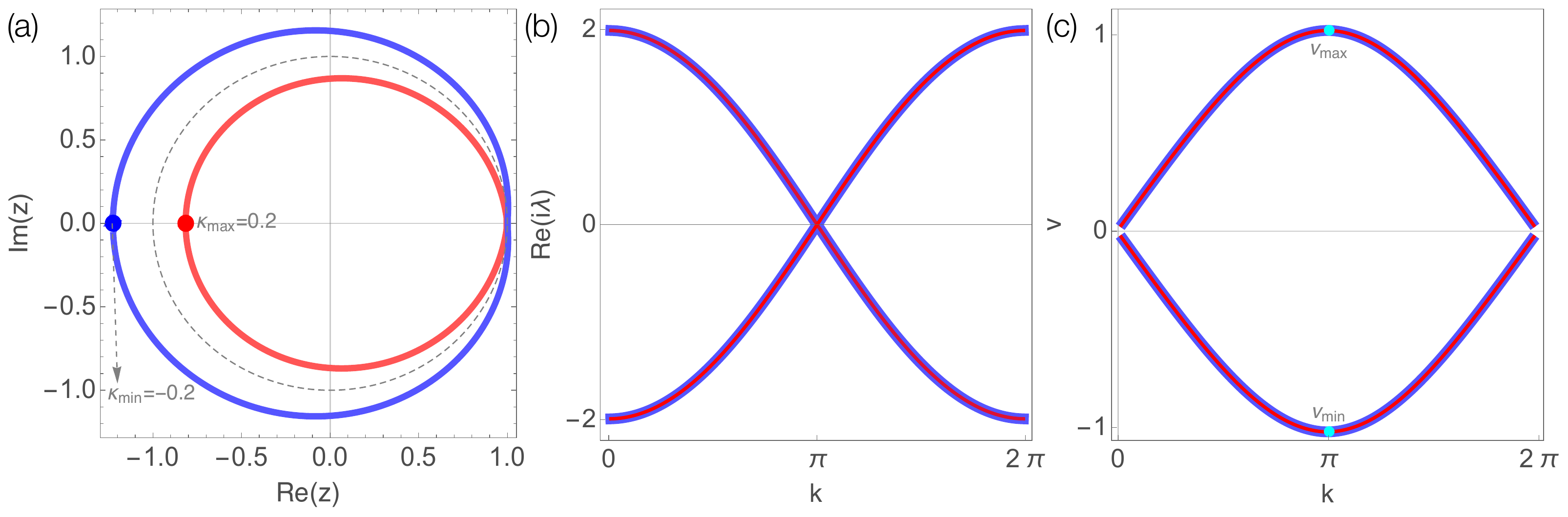}\\
\caption{(a) The GBZ of Eq.(\ref{xmatrix}). Parameter values are $t_1=t_2=1$, $\gamma=0.4$, $\delta_1=0.1$ and $\delta_2=0$. Two bands correspond to red curves and the other two to blue curves.
(b) $Re(i\lambda_{k+i\kappa})$ as a function of $k$ and band index, there are four bands and each band corresponds to one curve. There are two red curves and two blue curves. (c) Velocity $v=\frac{i\partial \lambda_{k+i\kappa,\alpha}}{\partial k}$ as a function of $k$ and band index. Curves in (a) (b) and (c) with the same color are corresponding each other. } \label{figsm5}
\end{figure}

For parameter values $t_1=t_2=1$, $\gamma=0.4$, $\delta_1=0.1$ and $\delta_2=0$, we illustrate the GBZ, $Re(i \lambda)$ versus $k$ and $v$ as a function of $k$ in Fig.\ref{figsm5} (a)-(c), respectively. We can get $v_{max}=max\left\{Re(\frac{i\partial \lambda_{k+i\kappa,\alpha}}{\partial k})\right\}\approx 1$, $v_{min}=min\left\{Re(\frac{i\partial \lambda_{k+i\kappa,\alpha}}{\partial k})\right\}\approx -1$, $\kappa_{max}= 0.2$,  and $\kappa_{min}= -0.2$ at $k=\pi$. There is an accidental symmetry $\sigma_z \tau_z(X+\frac{\gamma}{2})=-(X+\frac{\gamma}{2})\sigma_z \tau_z$, which protects $v_{max}=-v_{min}$.

  Furthermore, if $\delta_1,\delta_2,\frac{\gamma}{t_1} \ll 1$:
\begin{equation}
\begin{split}
X \approx & i\left[ \begin{array}{cc}
H_{nSSH}(k) +\frac{i\gamma}{2}& 0\\
0& H_{nSSH}^T(-k)+\frac{i\gamma}{2}
\end{array}
\right ] \\
=&(-\frac{\gamma}{2}+it_1\sigma_x+\frac{\gamma}{2}\sigma_y\tau_z)+it_2\sigma_y \sin k +it_2\sigma_x \cos k.  \label{xmatrix1}
\end{split}
\end{equation}
The eigenvalues of open boundary $X$ matrix are:
\begin{equation}
\lambda_{k+i\kappa,\alpha=1,2,3,4}=-\frac{\gamma}{2}\pm i\sqrt{t_1^2+t_2^2+2t_1t_2 \cos (k+i\kappa)-\frac{\gamma^2}{4}\pm i\gamma t_2 \sin (k+i\kappa)} .
\end{equation}
We can get $v_{max}=-v_{min}\approx min\left\{t_1,t_2\right\}$ and $\kappa_{max}=-\kappa_{min}\approx \frac{\gamma}{2t_1}$ at $k=\pi$. So helical damping still exists for $\delta_1,\delta_2,\frac{\gamma}{t_1} \ll 1$ and $t_1<t_2$.

\section{SV: Prove the equivalence of GBZ equations for case 1 and case 3}
For case 1, the GBZ equations are $|z_{1}^a|=|z_{2}^a|$ and $|z_{1}^b|=|z_{2}^b|$, where $|z_1^a|\le|z_2^a|$ and $z_{i}^b=\frac{1}{z_{i}^a}$  $(i=1,2)$. Without loss of generality, we assume that $|z_2^a|\le 1$. Then the roots of $f(z)=0$ are $z_1^a,z_2^a,z_1^b,z_2^b$ with $|z_1^a|\le|z_2^a|\le|z_2^b|\le|z_1^b|$, and GBZ equations are $|z_{1}^a|=|z_{2}^a|$ and $|z_{1}^b|=|z_{2}^b|$, where $z_{i}^b=\frac{1}{z_{i}^a}$. And it is the GBZ equation of case 3.

\end{document}